\documentclass[11pt,titlepage]{article}
\usepackage{graphicx}
\usepackage{setspace}
\usepackage[round,numbers,sort&compress]{natbib}
\usepackage[utf8]{inputenc}
\usepackage[a4paper]{geometry}
\usepackage{amsmath,amssymb}
\usepackage{color}
\usepackage{bm}

\renewcommand{\d}{\mathrm{d}}

\newcommand{\DEF}{\stackrel{\mbox{\rm\scriptsize def}}{=}}
\definecolor{orange}{rgb}{0.87,0.38,0.16}
\definecolor{purple}{rgb}{0.52,0.34,0.56}

\title{Elasticity and electrostatics of plectonemic DNA}

\author{%
N.~Clauvelin, B.~Audoly, and S.~Neukirch\thanks{
	Corresponding author.  Address: 
	Institut Jean le Rond d'Alembert,
	Universit\'e Pierre et Marie Curie (case 162),
	4, place Jussieu,
	Paris~F-75005, France,
	Tel.:~(+33)1 44 27 37 90} \\
UPMC Univ Paris 06, UMR 7190, Institut Jean Le Rond d'Alembert, F-75005 Paris, France.\\
CNRS, UMR 7190, Institut Jean Le Rond d'Alembert, F-75005 Paris, France.
}

\date{\today}
\begin{document}
\sloppy
\maketitle

\abstract{We present a self-contained theory for the mechanical
response of DNA in single molecule experiments.  Our model is based on
a 1D continuum description of the DNA molecule and accounts both for
its elasticity and for DNA-DNA electrostatic interactions.  We consider
the classical loading geometry used in experiments where one end of
the molecule is attached to a substrate and the other one is pulled by
a tensile force and twisted by a given number of turns.  We focus on
configurations relevant to the limit of a large number of turns, which
are made up of two phases, one with linear DNA and the other one with
superhelical DNA. The model takes into account thermal fluctuations in
the linear phase and electrostatic interactions in the superhelical
phase.  The values of the torsional stress, of the supercoiling radius
and angle, and key features of the experimental extension-rotation
curves, namely the slope of the linear region and thermal buckling
threshold, are predicted. They are found in good agreement with
experimental data.

 } 
\clearpage

\section{Introduction}

\begin{figure}[!ht]
	\centering
	\includegraphics[width=.705\columnwidth]{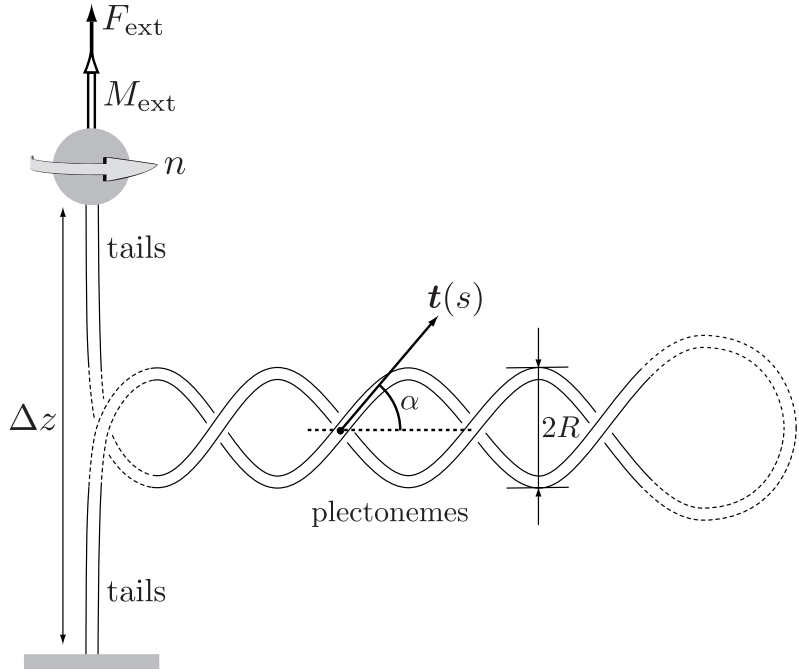}
	\caption{ Sketch of the experimental setup: a dsDNA molecule
	is fixed by one end to a glass pane while the other end is
	attached to a mechanical system, symbolized by the grey disc,
	which allows one to exert a pulling force $F_{\mathrm{ext}}$
	and impose a rotation $2\pi n$.  For large numbers of turns
	the molecule coils around itself in a helical way and forms
	plectonemes.  The configuration of the molecule is made of two
	phases: the tails and the plectonemes.  The plectonemic phase
	is characterized by superhelical radius $R$ and angle $\alpha$
	between the tangent $\bm{t}(s)$ and the
	helices axis.  The dashed parts represent the regions we
	neglect: the matching region between the tails and the
	plectonemes and the end loop.  }
	\label{fig:PlectonemsGeometry}
\end{figure}

Mechanics of the DNA molecule plays a key role in several biological
processes at the cellular level.  In several cases, the action of
enzymes and proteins on DNA has been found to depend on the mechanical
stress present in the molecule.  For instance, the torsional moment in
DNA controls the action of topoisomerases or
RNA-polymerases~\citep{Koster:Friction-and-torque-govern-the-relaxation-of-DNA-supercoils-by-eukaryotic-topoisomerase-IB:2005,
revyakin+al:2004}.  In this context, experiments where forces and torques
are applied to a single DNA molecule provide a remarkable opportunity
to gain insights into the mechanics of DNA. We are here interested in
extension-rotation experiments using either optical or magnetic
tweezers~\citep{smith+al:1992,
Strick:The-Elasticity-of-a-Single-Supercoiled-DNA-Molecule:1996,
Bustamant:Grabbing-the-cat-by-the-tail:-manipulating-molecules-one-by-one:2000,
Charvin:Twisting-DNA:-single-molecule-studies:2004,
Deufel:Nanofabricated-quartz-cylinders-for-angular-trapping:-DNA-supercoiling-torque-detection:2007}, see
Ref.~\citep{Ritort:Single-molecule-experiments-in-biological-physics:-methods-and-applications:2006}
for a review.
These experimental setups are equivalent from a mechanical
perspective: a dsDNA molecule is fixed at one end on a glass pane
while the other end is attached to a bead that pulls and twists it.
In these experiments, traction and rotation are controlled
differently: for the rotation mode, the twist angle is prescribed and
the twist moment varies accordingly; for the stretching mode, the
extension can vary although the pulling force is prescribed.  DNA is
under or over-wound and various molecule conformations are
observed~\citep{Charvin:Twisting-DNA:-single-molecule-studies:2004}.
In the present study, we focus on the over-winding of a dsDNA molecule
under large imposed rotations:
%
%
the molecule coils around itself in a helical
way and forms plectonemes, as sketched in
Fig.~\ref{fig:PlectonemsGeometry}.  
An important feature of the experimental
loading curves is the linear decrease of the vertical extension of the
molecule as a function of the imposed rotation.  We have shown in
previous
studies~\citep{Clauvelin:Mechanical-Response-of-Plectonemic-DNA:-An-Analytical-Solution:2008,
Neukirch:Extracting-DNA-Twist-Rigidity-from-Experimental-Supercoiling-Data:2004}
that this behavior can be captured by a purely elastic rod model based
on Kirchhoff-Love elastic rod theory.  In this paper, we extend this
model and investigate the combined effects of elasticity and
electrostatics.

The response of plectonemic DNA under stress involves various physical
phenomena such as elastic deformations, thermal fluctuations,
electrostatic interactions, self-avoidance.  Although some of these
effects have been considered in the literature, a model addressing
them all together is still lacking.  Mechanical models of twisted rods
in contact have been introduced, from an
analytical~\citep{Coleman:Theory-of-self-contact-in-Kirchhoff-rods-with-applications-to-supercoiling-of-knotted-and-unknotted-DNA-plasmids:2004}
or
numerical~\citep{Goyal:Nonlinear-dynamics-and-loop-formation-in-Kirchhoff-rods-with-implications-to-the-mechanics-of-DNA-and-cables:2005}
perspective, but thermal fluctuations are not treated.  A simplified
analytical model, including some account for fluctuations but omitting
contact forces in the plectonemic region, is proposed in
Ref.~\citep{Purohit:Plectoneme-formation-in-twisted-fluctuating-rods:2008}.
Statistical mechanics of plectonemic DNA has been approached, either
analytically~\citep {fain+rudnick:1997,bouchiat+mezard:2000} or
numerically~\citep {vologodskii+marko:1997} using a Monte-Carlo
method.  The validity of some of these results was questioned in
Ref.~\citep{rossetto+maggs:2003,Neukirch:Writhe-formulas-and-antipodal-points-in-plectonemic-DNA-conf:2008};
in addition, long-range potentials raise convergence issues that have
not yet been overcome in Monte-Carlo simulations.  A composite model,
gathering results from torsionally constrained
polymer~\citep{moroz+nelson:1998} and Monte-Carlo simulations has
recently been introduced~
\citep{Marko:Torque-and-dynamics-of-linking-number-relaxation-in-stretched-supercoiled-DNA:2007}.
To date, this is the only model that confronts its predictions against
experimental data.  However it relies on an assumption on the
supercoiling free energy that is not always
valid~\citep{Klenin:Computer-simulation-of-DNA-supercoiling:1991,Marko-Micromechanics-of-single-supercoiled-2008}
and uses parameters extracted from Monte-Carlo simulations.

In this paper, we present a self-contained analytical model for the
mechanical response of plectonemic DNA in extension-rotation
experiments, which builds up on previous
work~\citep{Clauvelin:Mechanical-Response-of-Plectonemic-DNA:-An-Analytical-Solution:2008}.
We focus on the plectonemic regime at large imposed rotations. This 
corresponds to the linear region in the 
experimental extension-rotation curves.  Our elastic model accounts for
DNA-DNA interactions in the plectonemic region and for thermal
fluctuations in the tail regions, where they are dominant. 
It captures the main features of the experimental curves and allows
quantitative comparison to experiments with no adjustable parameter.

The paper is organized as follows.  In the next section we present our
model and derive the equilibrium equations for a DNA molecule
comprising plectonemes, for a generic interaction potential.  In
section~\ref{section:dna-dna_interactions}, we describe two
representative DNA-DNA interactions potentials available in the
literature, which we then plug into our model.
The results are then compared with experimental
data in section~\ref{section:results}.

\section{Model}

Our description of the DNA molecule is based on a coarse-grained
representation~\cite{art:lebret}.  We introduce a continuum rod model
whose mechanical behavior is similar to that of the molecule, and
makes use of effective elastic and electrostatic properties obtained
by smoothing out the details at a scale of several base pairs.
We deal with  an inextensible elastic rod with 
circular cross section, bending rigidity $K_0$, and twisting rigidity
$K_3$.  
The loading geometry is that of Fig.~\ref{fig:PlectonemsGeometry}, and
applies to the experiments where the lower end of the molecule is
clamped on a glass pane and the other end is subjected to a tensile
force $F_{\mathrm{ext}}$ and rotated by $n$ turns (\emph{i.e.} an
angle $2\pi\,n$).  The imposed rotation is achieved through a
torsional moment $M_{\mathrm{ext}}$.
Note that the torsional moment has become accessible to experimental
measurements~\citep{Deufel:Nanofabricated-quartz-cylinders-for-angular-trapping:-DNA-supercoiling-torque-detection:2007}
only recently.

\subsection*{Geometry}

The inextensible rod, of length $\ell$, is parametrized by its
arc-length $s$, the origin $s=0$ being at the lower end.  The rod
centerline is described by a vector-valued function $\bm{r}(s)$ and
its unit tangent $\bm{t}(s) \DEF \mathrm{d}\bm{r}/\mathrm{d}s$.  The
geometric curvature of the rod is noted $\kappa(s) \DEF |
\mathrm{d}\bm{t}/\mathrm{d}s |$.  The twist is noted $\tau(s)$: it
describes the relative rotation of neighboring cross sections about
the tangent $\bm{t}(s)$.  Note that the twist is a different quantity
from the Fr\'enet (geometric) torsion of space curves --- the latter
is irrelevant in the context of elastic rods.

We consider the geometry of the double stranded DNA sketched in
Fig.~\ref{fig:PlectonemsGeometry}, which is relevant to the
plectonemic regime: two twisted, straight tails are separated by a
plectonemic region composed of two identical and uniform helices.
Note that each helix is itself a piece of the double stranded DNA
molecule.  For a large number of
turns $n$, the loop at the end of the plectonemes and the curved region
connecting the tails to the plectonemes are much smaller than the tails
and the helical parts, and hence are neglected.  Even though we depict the
plectonemic region as a single chunk for simplicity, our model applies
equally well to the case where the plectonemes are distributed in
several places along the molecule; then, the elastic rod is made up of
two \emph{phases}, one with linear DNA and the other one with
plectonemes, and the plectonemic structure in
Fig.~\ref{fig:PlectonemsGeometry} represents the plectonemic phase
collectively.  The molecule contour length spent in the tails phase
and in the plectonemes are noted $\ell_{\mathrm{t}}$ and $\ell_{\mathrm{p}}$
respectively.  They sum up to the total length $\ell=\ell_{\mathrm{p}}+\ell_{\mathrm{t}}$.
The plectonemic phase is characterized by its superhelical radius $R$
and its superhelical angle $\alpha$, which are assumed to be uniform:
neither $R$ nor $\alpha$ may depend on $s$, although they depend on
the loading.  Curvature is zero in the straight tails, and takes a
constant value in the plectonemes which can be evaluated using simple
geometry.  The integrated squared curvature, which enters into the
bending energy, is then found to be:
\begin{equation}
	\int_0^\ell \kappa^2(s)\,\d s =
	\frac{\sin^4\alpha}{R^2}\ell_{\mathrm{p}}\; ,
	\label{eqn:RodCurvature}
\end{equation}
see
Ref.~\citep{Clauvelin:Mechanical-Response-of-Plectonemic-DNA:-An-Analytical-Solution:2008}.
Since the rod has a circular cross section, the twist $\tau(s)$ is
uniform, $\mathrm{d}\tau/\mathrm{d}s=0$, as shown for instance in
Ref.~\cite{love:1944}.  The internal torsional moment $M(s)$ in the
rod is related to the twist $\tau(s)$ by the constitutive law
$M(s)=K_3 \, \tau(s)$.  Therefore, its value $M(s)$ is constant along
the rod, and equal to the torque $M_{\mathrm{ext}} = K_{3}\,\tau$
applied by the bead.  In what follows, we study the equilibrium of the
rod and compute the parameters $R$, $\alpha$, $\ell_{\mathrm{p}}$ and
$\tau = M_{\mathrm{ext}} / K_{3}$ as a function of the loading
(pulling force $F_{\mathrm{ext}}$ and number of turns $n$) by
minimizing the energy.

\subsection*{Variational formulation}

We derive the energy of the system as a function of the superhelical
angle $\alpha$ and radius $R$, of the twist $\tau$ and of the
plectonemic contour length $\ell_{\mathrm{p}}$.  
Equilibrium solutions and their stability will be
derived later on by minimizing this energy.
The experiments are performed under imposed end
rotation: energy minimization is performed under the constraint that
the number of turns $n$ imposed on the bead is equal to the link 
$\mathrm{Lk}$ of
the DNA molecule.  Neglecting the writhe of the
tails, the link can be written
as~\citep{Neukirch:Extracting-DNA-Twist-Rigidity-from-Experimental-Supercoiling-Data:2004,
Clauvelin:Mechanical-Response-of-Plectonemic-DNA:-An-Analytical-Solution:2008}:
\begin{equation}
	n=
    \mathrm{Lk}=\mathrm{Tw}+\mathrm{Wr}=
	\frac{1}{2\pi}\int^\ell_0 \tau\,\d s - \chi\frac{\sin 2\alpha}{4\pi R}=
	\frac{1}{2\pi}\left(\tau\,\ell-\chi\frac{\sin 2\alpha}{2R}\ell_{\mathrm{p}}\right)\;,
	\label{eqn:LinkEquation}
\end{equation}
where $\chi=\pm1$ stands for the chirality of the two helices of the
plectonemic phase.

The total energy of the system is the sum of three terms,
$V=V_{\mathrm{el}}+V_{\mathrm{ext}}+V_{\mathrm{int}}$, where the first is the strain
elastic energy, the second is the potential energy associated with the
external load $F_{\mathrm{ext}}$, and the last term accounts for
DNA-DNA interactions between the two helices in the plectonemic phase.
The strain elastic energy of the rod is defined as the sum of a
bending term, proportional to the integrated squared curvature, and a
stretching term, proportional to the integrated squared twist:
\begin{equation}
    V_{\mathrm{el}}
    =\frac{K_0}{2}\int_0^\ell\kappa^2\mathrm{d}s + \frac{K_3}{2}\int_0^\ell\tau^2\mathrm{d}s
    = \frac{K_0}{2}\frac{\sin^4\alpha}{R^2}\ell_{\mathrm{p}} + \frac{K_3}{2}\tau^2\ell
    \;.
    \label{eqn:ElasticEnergy}
\end{equation}
This energy captures the elastic behavior of the rod in response to
applied forces and moments; it is zero in the natural (straight,
twistless) configuration of the rod.  The pulling force is described
using a potential energy:
\begin{equation}
    V_{\mathrm{ext}}
    = -F_{\mathrm{ext}} \left( z(\ell)-z(0) \right) = 
    -F_{\mathrm{ext}} \,\Delta z\;.
    \label{eqn:ExternalWork}
\end{equation}
Here $\Delta z \DEF ( z(\ell)-z(0) )$ is the extension of the molecule
along the direction $z$ of application of the pulling force.  Since we
assume the tails to be straight and neglect the curved region region
connecting the tails and the plectonemes, the vertical extension of the
filament reads $\Delta z = \ell_{\mathrm{t}}= \ell-\ell_{\mathrm{p}}$ and we can rewrite:
\begin{equation}
	V_{\mathrm{ext}} = -F_{\mathrm{ext}} \left( \ell-\ell_{\mathrm{p}} \right)\;.
	\label{eqn:ExternalWorkEnergy}
\end{equation}
There is no need to consider any potential energy associated with the
rotation of the end attached to the bead since the energy will be
minimized for a given rotation of the bead using the constraint on the
link.

In previous
work~\cite{Clauvelin:Mechanical-Response-of-Plectonemic-DNA:-An-Analytical-Solution:2008}
we solved this elastic rod model by assuming the superhelical radius
to be prescribed and extracted its value from experimental data.
Here, we take a more principled approach and complement the above
elastic equations with a proper model for DNA-DNA interactions in the
plectonemes; in particular, this makes it possible to predict the
superhelical radius.  These interactions are dominated by different
physical effects depending on the separation distance between the two
DNA superhelices.  In the range of separations relevant to
extension-rotation experiments, of order several nanometers,
electrostatic effects dominate.  In our model, interactions are
limited to the plectonemic phase and are described by an energy
contribution of the form:
\begin{equation}
	V_{\mathrm{int}}= \ell_{\mathrm{p}}\,U\!(R,\,\alpha)\;.
	\label{eqn:InteractionEnergy}
\end{equation}
This energy depends on the superhelical parameters $R$ and $\alpha$,
and is proportional to the plectonemic contour length
$\ell_{\mathrm{p}}$, and assumption valid when $\ell_{\mathrm{p}}$ is
much larger than $R$.

The total energy of the system is defined as the sum of the elastic, 
potential and interaction contributions:
\begin{equation}
    V(R,\alpha,\ell_{\mathrm{p}},\tau) = \frac{K_0}{2}\frac{\sin^4\alpha}{R^2}\ell_{\mathrm{p}} + \frac{K_3}{2}\tau^2\ell
    -F_{\mathrm{ext}} \left( \ell-\ell_{\mathrm{p}} \right)
    +\ell_{\mathrm{p}}\,U\!(R,\,\alpha)\,.
    \label{eqn:TotalEnergy}
\end{equation}
It will be minimized subjected to the end rotation constraint
given by Eq.~\ref{eqn:LinkEquation}.  This constraint provides an 
affine relation between $n$ and $\ell_{\mathrm{p}}$
and so can be used to eliminate the quantity $\ell_{\mathrm{p}}$ in favor of
$n$.  Dropping the constant term $(-F_{\mathrm{ext}}\,\ell)$ in the
energy, we obtain:
\begin{equation}
    V(\alpha, R, \tau) = \frac{K_3}{2}\,\tau^2\,\ell +
    \left(  2\pi n - \tau\,\ell \right)
    \left[
    \frac{-2 \chi}{\sin 2\alpha}
    \left(
    \frac{K_0}{2}\,\frac{\sin^4\alpha}{R} +R\,F_{\mathrm{ext}} +R\,U\!(R,\,\alpha)
    \right)
    \right]\,. \label{eqn:ConstrainedEnergy}
\end{equation}

\subsection*{Equilibrium equations}

The total energy of the system, given by
Eq.~\ref{eqn:ConstrainedEnergy}, takes into account the fixed end
rotation since Eq.~\ref{eqn:LinkEquation} has been used to eliminate
$\ell_{\mathrm{p}}$.  The equilibria of the rod are then directly given by
minimization of $V(\alpha, R, \tau)$ with respect to its three
arguments. Canceling the first variation of $V$, that is writing 
$\frac{\partial V}{\partial\alpha}=0$, 
$\frac{\partial V}{\partial R}=0$ and
$\frac{\partial V}{\partial\tau}=0$, we obtain
\begin{subequations}
    \label{eqn:EquilibriumEquations}
    \begin{align}
	2 K_0 \frac{\cos\alpha\,\sin^3\alpha}{R^2} + \frac{\partial 
	U(R,\alpha)}{\partial \alpha} - \frac{2}{\tan2\alpha}
	\left(\frac{K_0}{2}\frac{\sin^4\alpha}{R^2}+F_{\mathrm{ext}}+U\!(R,\,\alpha)\right) &=0\;,
	\label{eqn:AlphaCondition}\\
	F_{\mathrm{ext}} - \frac{K_0}{2\,R^2}\sin^4\alpha + R\,\frac{\partial U(R,\alpha)}{\partial R}+U\!(R,\,\alpha) &=0\;,
	\label{eqn:RCondition}\\
	M_{\mathrm{ext}} + \frac{2\chi}{\sin2\alpha}
	\left(\frac{K_0}{2}\frac{\sin^4\alpha}{R}+R\,F_{\mathrm{ext}}+R\,U\!(R,\,\alpha)\right) &=0
	\label{eqn:TauCondition}\;.
    \end{align}
\end{subequations}
In the first term of the last equation, we have eliminated $\tau$ in
favor of $M_{\mathrm{ext}}$ using the constitutive relation
$M_{\mathrm{ext}}=K_3\,\tau$, and thereby removed the twist rigidity
from the equations (its value is not known with good accuracy).

The set of three nonlinear equations~(\ref{eqn:EquilibriumEquations})
must be solved for the three unknown values of the parameters
$\alpha$, $R$ and $M_{\mathrm{ext}}$ at equilibrium, given the value
of the external force $F_{\mathrm{ext}}$.  This requires an
interaction potential $U\!(R,\,\alpha)$ to be specified, as is done in
the forthcoming Section~\ref{section:dna-dna_interactions}.  The set
of equations~(\ref{eqn:EquilibriumEquations}) extends the model of
Ref.~\citep{Clauvelin:Mechanical-Response-of-Plectonemic-DNA:-An-Analytical-Solution:2008},
valid for non-penetrable tubes, to filaments in long-range interaction
(such as electrostatic interactions).

Note that the equations~(\ref{eqn:EquilibriumEquations}) do not depend
on the number of turns $n$.  As a result, their solution $\alpha$,
$R$, $M_{\mathrm{ext}}$ do not depend on $n$ either.  The equations
describes the equilibrium of two phases; increasing $n$ transfers some
arc length from the tail phase to the plectonemic phase, without
changing their properties.  This invariance with respect to $n$
explains the presence of a linear region in the experimental curves,
as shown in the next section.

The term in parenthesis in Eq.~\ref{eqn:TauCondition} is always
positive. This shows that the sign of the chirality $\chi = \pm 1$
is opposite to that of $n$: rotating the bead in the positive
direction $n>0$ for instance, requires a positive torque
$M_{\mathrm{ext}}$, hence a negative $\chi = -1$ by this equation 
(left-handed superhelices).

\subsection*{Vertical extension of the filament}
%
In extension-rotation experiments the vertical extension of the
filament is recorded while the number of turns is increased.  The
formula $\Delta z = \ell_{\mathrm{t}}$, valid for straight tails, does
not holds in the presence of thermal fluctuations.
We account for these fluctuations by introducing a rescaled quantity:
\begin{equation}
    \Delta z_{\mathrm{th}} = \rho_{\mathrm{wlc}}\,\Delta z \, ,
    \label{equa:deltaz-thermique}
\end{equation}
where the correcting factor $\rho_{\mathrm{wlc}}$ is given by the
worm-like chain theory~\citep{Marko:Stretching-DNA:1995} as the
solution of:
\begin{equation}
    \frac{F_{\mathrm{ext}} K_0}{(k_B T)^2} 
    = \rho_{\mathrm{wlc}} + \frac{1}{4} 
    \frac{1}{(1-\rho_{\mathrm{wlc}})^2}-\frac{1}{4}\,.
\end{equation}
Here $k_B$ is the Boltzmann constant and $T$ the absolute
temperature.
In order to write $ \Delta z_{\mathrm{th}}$ as a function of the
number of turns $n$, we use the equality $\Delta
z=\ell_{\mathrm{t}}=\ell - \ell_{\mathrm{p}}$ in the right-hand side
of Eq.  \ref{equa:deltaz-thermique}, and use for $\ell_p$ the
expression obtained by solving Eq.~\ref{eqn:LinkEquation}:
\begin{equation}
    \Delta z_{\mathrm{th}} = 
    \left( 1 - \chi\frac{2R}{\sin2\alpha}\tau\right)\rho_{\mathrm{wlc}}\,\ell 
    + \chi\,\rho_{\mathrm{wlc}}\,\frac{4\pi R}{\sin2\alpha}\,n
    \label{eqn:VerticalExtension}\;.
\end{equation}
Recall that neither $\alpha$, $R$ nor $\tau = M_{\mathrm{ext}}/K_{3}$
depend on $n$; as a result, the extension $\Delta z_{\mathrm{th}}$
depends linearly on the number of turns $n$ in the above equation.
This linear dependence is a well-known feature of the experimental
curves.

\section{DNA-DNA interactions}
\label{section:dna-dna_interactions}

In the variational formulation exposed in the previous section we have
introduced an energy $U\!(R,\,\alpha)$ describing DNA-DNA
interactions.  At moderate distances DNA-DNA interactions in solution
mainly originate from electrostatic effects between the charged sites
of the two molecules (phosphate groups) and between these charged
sites and the counter and co-ions present in the solution.  The
theoretical analysis of the long and short range interactions between
two poly-ions in solution has been the subject of numerous
studies~\cite{Brenner:A-Physical-Method-for-Der:1974,Podgornik:Molecular-fluctuations-in:1990}
and there is currently a regain of interest in this question due to
recent progress in single-molecule experiments --- see
\citep{Kornyshev:Structure-and-interactions-of-biological-helices:2007}
for a review.  In the present model the interaction energy
$U\!(R,\,\alpha)$ is specified independently of the mechanical
behavior of the molecule.  As a result, we can combine the elastic
description of the previous sections with different theories for
DNA-DNA interactions.  In the following, we illustrate this approach
using two representative interaction energies $U(\alpha,R)$ that can
be found in the literature.

We favor interaction energies $U(R,\alpha)$ that can be expressed in
closed analytical form and have no adjustable parameters; this enables
us to make predictions and compare them to experiments, rather than to
fit existing data.  In the literature on DNA-DNA
interactions~\citep{Oosawa:Interaction-between-parallel-rodlike-macroions:1968,
Kornyshev:Structure-and-interactions-of-biological-helices:2007,
Tellez:Exact-asymptotic-expansions-for-the-cylindrical-Poissonndash;Boltzmann-equation:2006},
we picked two well-established models satisfying those
requirements.  The first one, $U_{PB}(R,\alpha)$,
derives from the Poisson-Boltzmann equation and was obtained by Ubbink
and
Odijk~\citep{Ubbink:Electrostatic-Undulatory-Theory-of-Plectonemically-Supercoiled-DNA:1999};
the second one, $U_{CC}(R,\alpha)$, is based on the counterion
condensation
theory~\citep{Manning:Limiting-Laws-and-Counterion-Condensation-in-Polyelectrolyte-Solutions-I.-Colligative-Properties:1969}
and was derived by Ray and
Manning~\citep{Ray:An-attractive-force-between-two-rodlike-polyions-mediated-by-the-sharing-of-condensed-counterions:1994}.
These two models address the electrostatics of DNA in solution but
their treatment of the interactions between DNA and the ions in
solution differ substantially.

\subsubsection*{Poisson-Boltzmann model}

In their study of supercoiled DNA
plasmids~\citep{Ubbink:Electrostatic-Undulatory-Theory-of-Plectonemically-Supercoiled-DNA:1999},
Ubbink and Odijk derive an analytical expression for the electrostatic
interaction energy between two interwound DNA molecules.  Their work
is based on the Poisson-Boltzmann framework (PB); in the computation
of the electrostatic repulsion of the two charged molecules, the
presence of the counter-ions and co-ions in solution is considered.
It has been shown in
Ref.~\citep{Stigter:Interactions-of-highly-charged-colloidal-cylinders-with-applications-to-double-stranded-DNA:1977}
that the non-linear PB problem could be simplified to a linear one by
considering screened (Debye-Huckel like) potentials and renormalized
linear charge densities $\nu$.  The value of the effective charge
$\nu$ is obtained by matching the solution of the non-linear PB
equation with the solution of the linear PB equation in the far-field
region.  However there is no consensus on the exact value of this
effective charge and the various
numerical~\citep{Marko:Statistical-mechanics-of-supercoiled-DNA:1995,
Vologodsk:Modeling-of-long-range-electrostatic-interactions-in-DNA:1995}
or
analytical~\citep{Stroobant:Effect-of-electrostatic-interaction-on-the-liquid-crystal-phase-transition-in-solutions-of-rodlike-polyelectrolytes:1986,
Trizac:Simple-Approach-for-Charge-Renormalization-in-Highly-Charged-Macroions:2002}
studies yield scattered results.
 
In
Ref.~\citep{Ubbink:Electrostatic-Undulatory-Theory-of-Plectonemically-Supercoiled-DNA:1999},
Ubbink and Odijk compute the electrostatic interaction energy per unit
length as
\begin{subequations}
    \label{eqn:OdijkEnergy}
    \begin{equation}
	U_{PB}(R,\alpha) = 
	\frac{1}{2}\, k_BT\, \nu^2\,l_B\,\sqrt{\frac{\pi}{\kappa_{D} 
	R}}\,e^{-(2\kappa_{D} R)} \, \varphi(\alpha)
	\;,
	\label{eqn:OdijkEnergy-U} 
    \end{equation}
where the angular dependence reads
\begin{equation}
	\varphi(\alpha) =1+0.83\,\tan^2\alpha + 0.86 \, \tan^4\alpha\;.
	\label{eqn:OdijkEnergy-phi}	
\end{equation}
\end{subequations}
Here, $k_B$ is the Boltzmann constant, $T$ the temperature in Kelvin,
$\nu$ the effective linear charge density (in m$^{-1}$), $l_B$ the
Bjerrum length, and $\kappa_{D}^{-1}$ the Debye length.  For a typical
temperature $T=300~\mathrm{K}$ we have $l_B=0.7~\mathrm{nm}$, and for
a monovalent salt concentration $c=10~\mathrm{mM}$ the Debye length is
$\kappa_{D}^{-1}=3.07~\mathrm{nm}$.  The value of the effective charge
$\nu$ depends on salt concentration, its value for a monovalent salt
concentration $c=10~\mathrm{mM}$ is taken as
$\nu=1.97~\mathrm{nm}^{-1}$ according to
Ref.~\citep{Ubbink:Electrostatic-Undulatory-Theory-of-Plectonemically-Supercoiled-DNA:1999}.

The calculation of the interaction energy can be simplified by taking
$\alpha = 0$, hence $\varphi(\alpha)=\varphi(0) = 1$, which amounts to
consider two straight and parallel molecules; this approximation has
been used for instance in
Ref.~\citep{Marko:Statistical-mechanics-of-supercoiled-DNA:1995}.  In
the rest of the paper $U_{PB}^{\dag}$ will refer to the potential
obtained under this approximation, namely $U_{PB}^{\dag}(R) =
U_{PB}(R,0)$.

\subsubsection*{Ray and Manning model}

The interaction energy derived by Ray and Manning
in~\citep{Ray:An-attractive-force-between-two-rodlike-polyions-mediated-by-the-sharing-of-condensed-counterions:1994}
is based on the counter-ions condensation
theory~\citep{Manning:Limiting-Laws-and-Counterion-Condensation-in-Polyelectrolyte-Solutions-I.-Colligative-Properties:1969}.
The authors examine the interaction of two straight and parallel DNA
molecules with charged sites in solution (the dependence on the
superhelical angle $\alpha$ is neglected).  The main point of the
theory is to consider that part of the DNA bare charge is neutralized
by the condensation of the counter-ions around the molecule.  The
energy is the sum of three terms: interactions between pairs of
charged sites belonging to the same DNA segment, interactions between
pairs of charged sides located on opposite segments, and adsorption
energy of the counter-ions onto the molecule.  Three cases are
considered, namely short, intermediate, and long interaxial distances
between the molecules.  The short distance case, below the
crystallographic radius of DNA, is not relevant to our analysis.  The
intermediate case introduces an adjustable parameter, which we try to
avoid.  Consequently we only use the long-range case,
relevant for inter-distances larger than the Debye length; in our
notations it writes:
\begin{equation}
	U_{CC}(R) = \frac{k_BT}{2 b} \,
	\left(2-\frac{1}{\xi}\right)\,B^0_K(2\kappa_{D} R)\;,
\end{equation}
where $b=0.17~\mathrm{nm}$ is the charge spacing parameter of the DNA
molecule, and $\xi=l_B/b$ is the dimensionless charge density of DNA
($\xi=4.11$ at $T=300~\mathrm{K}$).  The function $B^0_K(x)$ is the
modified Bessel function of the second kind and order 0.

\section{Results}
\label{section:results}

We solve equations~\ref{eqn:EquilibriumEquations} for the superhelical
radius $R$, angle $\alpha$, and external torque $M_{\mathrm{ext}}$,
using one of the interaction energies $U_{PB}$, $U_{PB}^{\dag}$ or
$U_{CC}$.  These equations are nonlinear and their roots are found
numerically using a Newton-Raphson algorithm.  We present the results
for the superhelical variables $R$ and $\alpha$ in
Fig.~\ref{fig:SuperhelicalDimensions}, for the torsional moment
$M_{\mathrm{ext}}$ in Fig.~\ref{fig:TorsionalMoment}.  We also plot
derived quantities, to be defined later, such as the slope $q$ of the
extension-rotation curves in Fig.~\ref{fig:PlectonemicSlope}, and the
thermal buckling threshold $n^{\star}$ in Fig.~\ref{fig:nstar_de_F}.
We compare our results with the model of
Ref.~\citep{Marko:Torque-and-dynamics-of-linking-number-relaxation-in-stretched-supercoiled-DNA:2007}
and with experimental data.  To ease comparison with our previous
work~\citep{Clauvelin:Mechanical-Response-of-Plectonemic-DNA:-An-Analytical-Solution:2008},
we use the same set of experimental data.  These data were obtained on
a 48 kbp lambda phage DNA molecule in a 10 mM phosphate buffer.

With the interaction energies used in the present paper, we find that
the nonlinear equations have two roots below a threshold value of the
force, and no root above.  For a salt concentration
$c=10~\mathrm{mM}$, this threshold value of the force is $
4.7~\mathrm{pN}$ using $U_{PB}$, $4.9~\mathrm{pN}$ using
$U_{PB}^\dag$, and $6.9~\mathrm{pN}$ using $U_{CC}$; all these values
are above the maximum pulling force applied in typical experiments.
We have studied the stability of the two solutions corresponding to
the two roots of our equations when the force is below threshold,
and found that that with lower $\alpha$ and $R$ is unstable; the other
one is stable.  We study and plot the latter in the following.

When the force reaches its threshold value, an instability occurs and
the stable solution disappears by merging with the unstable one.  For
larger forces, no stable solution exist and the two helical parts
collapse.  This may be related to the observation of tightly
supercoiled configurations in Ref.~\cite{bednar+al:1994}.  The
collapse arises when the electrostatic interaction is no longer strong
enough to sustain the applied force; the description of collapsed
solutions would require an account of hard-core repulsion and other
short-range forces.

\subsubsection*{Superhelical geometry}

\begin{figure}[tb]
	\centering
	\includegraphics[width=.7\columnwidth]{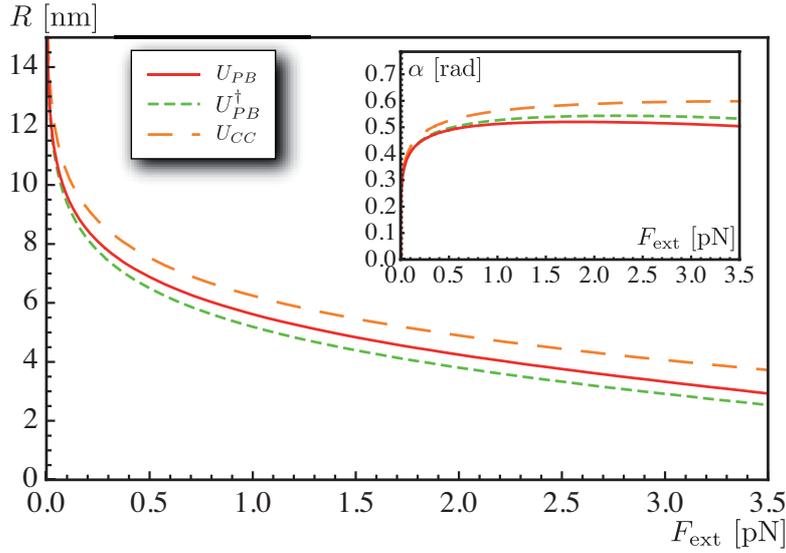}
	\caption{Computed values of the superhelical radius $R$ and
	angle $\alpha$ (inset) as functions of the pulling force
	$F_{\mathrm{ext}}$, using one of the interaction energies
	$U_{PB}$, $U_{PB}^{\dag}$ or $U_{CC}$.  These plots are
	obtained by solving the equilibrium
	equations~\ref{eqn:EquilibriumEquations} for each value of the
	pulling force $F_{\mathrm{ext}}$.  The function
	$\alpha(F_{\textrm{ext}})$ decreases for all interaction
	energies at large enough forces --- with the energy $U_{CC}$, this
	decrease occurs beyond the domain of forces shown in the
	figure.}
	\label{fig:SuperhelicalDimensions}
\end{figure}

The quantities $R$ and $\alpha$ are plotted 
Fig.~\ref{fig:SuperhelicalDimensions} as a function of the applied 
force $F_{\mathrm{ext}}$.
The curves $R(F_{\textrm{ext}})$ and $\alpha(F_{\textrm{ext}})$
obtained for the different interaction energies $U_{PB}$,
$U_{PB}^{\dag}$ or $U_{CC}$ are close over the entire range of forces.
As will be confirmed later, the predictions based on the different
interaction models are all very similar.

As expected, the superhelical radius decreases with the pulling force;
note that it becomes less than the Debye length for forces above
$F_{\textrm{ext}} \approx 2.5~\mathrm{pN}$.  The superhelical angle
$\alpha$ is known to be a control parameter in the action of the
topoisomerases~\cite{Neuman:Untwisting-and-untangling-DNA:-Symmetry-breaking-by-topoisom:2008}.  It is plotted
in the inset of Fig.~\ref{fig:SuperhelicalDimensions}.  In contrast
with models of elastic tubes in
contact~\cite{Neukirch-Heijden:pipotage}, where $\alpha$ increases
monotonically and reaches the value $\pi/4$ asymptotically at large
forces, we find here that it reaches a maximum well below $\pi/4$ and
then decreases, due to long-range forces.  This decrease has already
been observed in
Ref.~\citep{Clauvelin:Mechanical-Response-of-Plectonemic-DNA:-An-Analytical-Solution:2008},
where the value of the superhelical angle was extracted from
experimental data.

\subsubsection*{Torsional moment}

\begin{figure}[tb]
	\centering
	\includegraphics[width=.7\columnwidth]{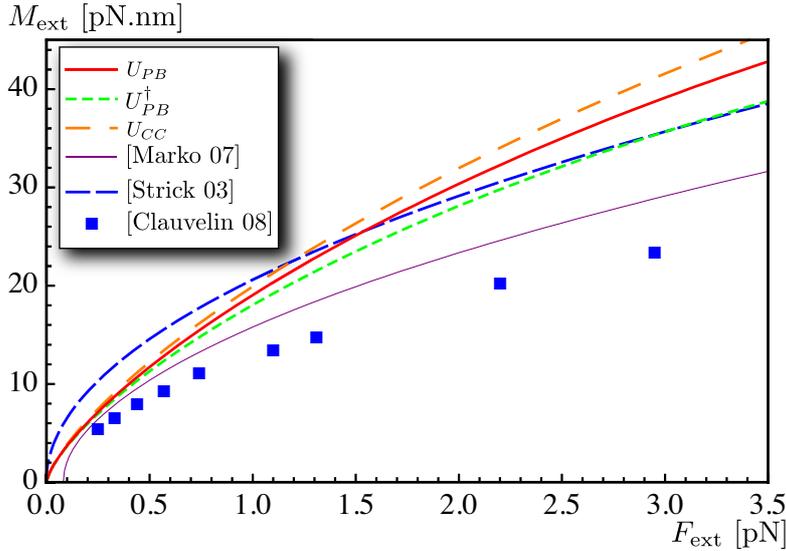}
\caption{ Computed values of the torsional moment in the molecule
$M_{\mathrm{ext}}$ as a function of the pulling force
$F_{\mathrm{ext}}$.  We compare the results of the present model using
the interaction potentials $U_{PB}$, $U_{PB}^{\dag}$ or $U_{CC}$, to
predictions in
Ref.~\citep{Marko:Torque-and-dynamics-of-linking-number-relaxation-in-stretched-supercoiled-DNA:2007}, namely
Marko's Eq.~17 with $A=50~\mathrm{nm}$, $C=95~\mathrm{nm}$ and
$P=28~\mathrm{nm}$, and Strick et
al.~\cite{Strick:Stretching-of-macromolecu:2003}, and results from our
previous
work~\citep{Clauvelin:Mechanical-Response-of-Plectonemic-DNA:-An-Analytical-Solution:2008}.
}
	\label{fig:TorsionalMoment}
\end{figure}

Recall that the torque $M_{\mathrm{ext}} = K_{3}\,\tau$ applied by the
bead in order to impose a rotation $2\pi\,n$ does not depend on the
number of turns $n$ by equations.~\ref{eqn:EquilibriumEquations}.  This quantity is plotted in
Fig.~\ref{fig:TorsionalMoment} as a function of the pulling force
$F_{\mathrm{ext}}$.  We compare our results (\emph{i}) with Marko's
Eq.~17 in
Ref.~\citep{Marko:Torque-and-dynamics-of-linking-number-relaxation-in-stretched-supercoiled-DNA:2007};
(\textit{ii}) with a formula $M_{\textrm{ext}} =
\sqrt{2\,K_{0}\,F_{\textrm{ext}}}$ obtained by Strick et al.  by
approximating the plectonemes by a chain of
circles~\cite{Strick:Stretching-of-macromolecu:2003}; and (\textit{iii})
with our previous
study~\citep{Clauvelin:Mechanical-Response-of-Plectonemic-DNA:-An-Analytical-Solution:2008}
based on hard-wall interactions.
Even though
Refs.~\citep{Marko:Torque-and-dynamics-of-linking-number-relaxation-in-stretched-supercoiled-DNA:2007},
\citep{Strick:Stretching-of-macromolecu:2003}
and~\citep{Clauvelin:Mechanical-Response-of-Plectonemic-DNA:-An-Analytical-Solution:2008}
do not address long-range interactions, all the curves reveal a
similar behavior: the moment increases monotonically with the applied
force, with a decreasing slope.  However, our results show that
long-range interactions significantly increase the value of the moment
required to achieve a given rotation.

\subsubsection*{Extension-rotation curve}

\begin{figure}[tb]
	\centering
	\includegraphics[width=.7\columnwidth]{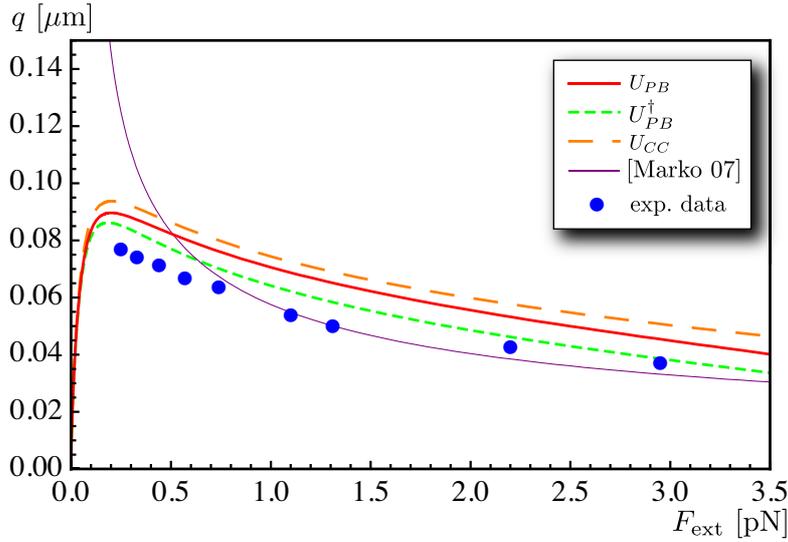}
	\caption{
	Computed values of the slope of the plectonemic region $q$ as a function of
the pulling force $F_{\mathrm{ext}}$. Experimental points with filled
circles are extracted from the experimental hat curves shown in Fig.~2 of
\cite{Clauvelin:Mechanical-Response-of-Plectonemic-DNA:-An-Analytical-Solution:2008}.
The thin purple curve is obtained from the theory
in~\citep{Marko:Torque-and-dynamics-of-linking-number-relaxation-in-stretched-supercoiled-DNA:2007},
using the same parameter values as in Fig.~\ref{fig:TorsionalMoment}.
	}
	\label{fig:PlectonemicSlope}
\end{figure}

\begin{figure}[!ht]
	\centering
	\includegraphics[width=.7\columnwidth]{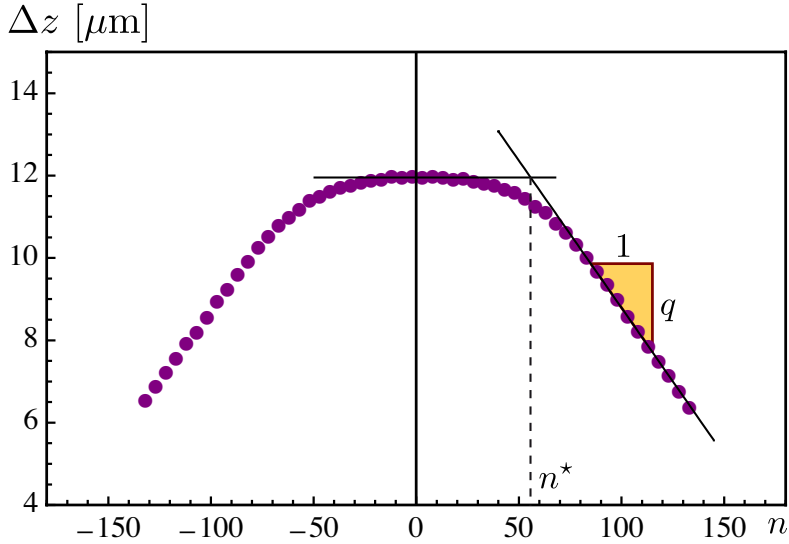}
\caption{ Experimental curve showing the vertical extension $\Delta
z_\mathrm{exp}$ of a lambda phage DNA 48kbp molecule as a function of
the imposed number of turns $n$, at constant force
$F_\mathrm{ext}=0.44$pN. The quantity $\Delta z_{\mathrm{th}}(n)$
defined in Eq.~\ref{eqn:VerticalExtension} is our prediction for the
linear part of the experimental curve.  The number of turns at the
transition $n^\star$ and the slope $q$ are also shown.}
	\label{fig:def_n_flamb}
\end{figure}

Our model predicts that the derivative of the vertical extension
$\Delta z_{\textrm{th}}$ with respect to $n$ is constant, \emph{i.
e.} that the extension-rotation curve is linear in the regime of large
rotations that we consider.  This linear regime is well-known
experimentally, see Fig.~\ref{fig:def_n_flamb}. From Eq.~\ref{eqn:VerticalExtension}, the slope $q$ is given by:
\begin{equation}
    q=\frac{\d \Delta z_{\mathrm{th}}}{\d n}=\chi\frac{4\pi R}{\sin2\alpha} \rho_{\mathrm{wlc}}\;.
    \label{eqn:PlectonemicSlope}
\end{equation}
Its value is computed using the values of $\alpha$ and $R$ obtained
earlier by solving the equilibrium equations.  We plot in
Fig.~\ref{fig:PlectonemicSlope} the slope $q$ as a function of the
force, for the three interaction energies.
For comparison, we also plot the slope predicted by Marko's
model~\citep{Marko:Torque-and-dynamics-of-linking-number-relaxation-in-stretched-supercoiled-DNA:2007},
and the slopes read off directly from experimental extension-rotation
curves (these experimental data were kindly provided by V. Croquette
and have appeared in Fig.~2 of our
Ref.~\citep{Clauvelin:Mechanical-Response-of-Plectonemic-DNA:-An-Analytical-Solution:2008}
and in Refs.~\citep{Neukirch:Extracting-DNA-Twist-Rigidity-from-Experimental-Supercoiling-Data:2004} and~\citep{Marko:Torque-and-dynamics-of-linking-number-relaxation-in-stretched-supercoiled-DNA:2007}).

Our model shows good agreement with the experimental data which are 
reproduced
in a more consistent manner, especially at low forces, than in
Ref.~\citep{Marko:Torque-and-dynamics-of-linking-number-relaxation-in-stretched-supercoiled-DNA:2007}.
In this reference, hard-wall interactions are used with a supercoiling radius
independent of the pulling force; this may be the cause of the poorer
agreement with experimental data at low forces, when long-range interactions
dominate.

\begin{figure}[!ht]
	\centering
	\includegraphics[width=.7\columnwidth]{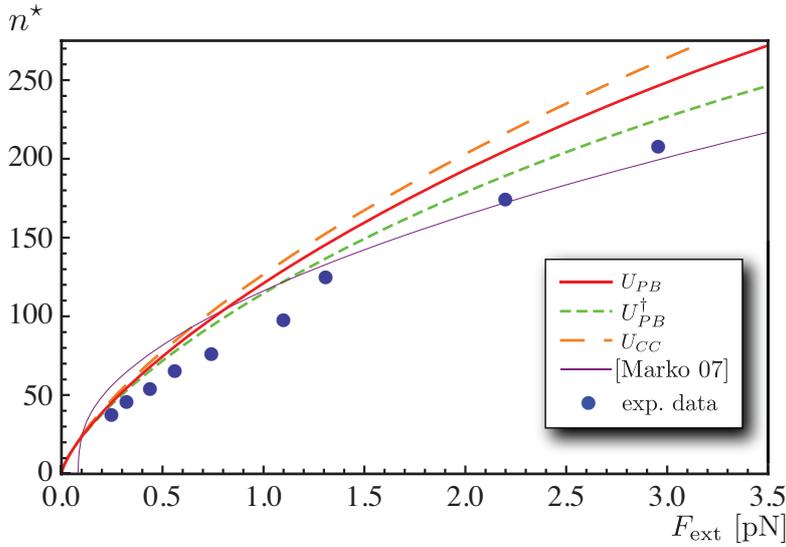}
	\caption{Computed values of the number of turns $n^\star$ at the transition as a function of the pulling force
	$F_{\mathrm{ext}}$, using $K_3=95$ nm $k_BT$. Experimental points are extracted from the curves shown in Fig.~2 of
	\cite{Clauvelin:Mechanical-Response-of-Plectonemic-DNA:-An-Analytical-Solution:2008}.
	The thin purple curve is obtained from the theory
	in~\citep{Marko:Torque-and-dynamics-of-linking-number-relaxation-in-stretched-supercoiled-DNA:2007},
	using the same parameter values as in Fig.~\ref{fig:TorsionalMoment}.
	}
	\label{fig:nstar_de_F}
\end{figure}

A typical extension-rotation curve comprises two regions: a linear region for
large $n$, which we have been discussing so far, and a parabolic region at low $n$ studied in
Ref.~\citep{moroz+nelson:1997}.
The central region is dominated by thermal effects and will not be
addressed here.  However, we can characterize the transition between
the two regions.  The number $n^\star$ of turns at which the
transition occurs is defined using the linear extrapolations shown in
Fig.~\ref{fig:def_n_flamb}.  This $n^\star$ corresponds to the onset of the
plectonemic regime. In our model, it is computed by setting
$\ell_\mathrm{p}=0$ in Eq.~\ref{eqn:LinkEquation}; this yields
$n^\star=\tau \ell/(2 \pi)$.  Recall that the value of
$\tau=M_{\mathrm{ext}}/K_3$ is computed from the equilibrium
equations~\ref{eqn:EquilibriumEquations}.  We plot in
Fig.~\ref{fig:nstar_de_F} the value of $n^\star$ as a function of the
force and compare to the values extracted from the experimental curve
as well as the value from the theory in
Ref.~\citep{Marko:Torque-and-dynamics-of-linking-number-relaxation-in-stretched-supercoiled-DNA:2007}.

\section{Discussion and conclusion} 
\label{section:conclusion}

In Section~\ref{section:results}, we found that the solutions
disappear above a threshold value of the pulling force.  This can be
interpreted as the fact that the physical solution involves a very
short inter-distance, although we have retained the long-distance part
of Ray and Manning's potential only.  This can be cured in principle
by restoring a complete expression~\footnote{Except in the
long-distance part used here, Ray and Manning's potential involves
physical quantities that are unknown.} of the potential given in
Ref.~\citep{Ray:An-attractive-force-between-two-rodlike-polyions-mediated-by-the-sharing-of-condensed-counterions:1994}.
In its complete form, the potential is non-monotonous and several
inter-distances are possible for a given value of the control
parameter.  This feature indicates the possibility of a transition
from a classical supercoiled state to a tight supercoiled state.  A
possibly related transition has been reported in
experiments~\cite{bednar+al:1994}.  When used in conjunction with the
complete Ray and Manning's potential, our model could provide a bridge
between the analytical expression for the potential and the
experiments, and provide a quantitative account of the transition.

We have presented an analytical model for DNA supercoiling in
extension-rotation experiments.  It is based on an elastic description
of DNA deformations, carefully accounts for DNA-DNA interactions in
the plectonemic region, and makes use of a valid formula for the link.
DNA interactions are modelled using long-range forces computed from
potentials available from the literature.  Their description is
compartmentalized from the rest of the theory, which makes it possible
to test different interaction energies.  We have used our model in
combination with two interaction energies.  These energies come from
different physical contexts and are widely used in the literature.
Using either one, we find good agreement with experimental data
without adjusting any parameter.  This suggests (\emph{i}) that using yet
another energy for electrostatic interactions would yield comparable
results and (\textit{ii}) that DNA tertiary structures are determined to a
large extent by the elasticity of the molecule and do not depend
heavily on the details of the interaction model.  The
extension-rotation experiments that are now routinely performed can
then be viewed as a way to probe the elastic properties of the
molecule.  Given that the mechanics of DNA under combined twist and
tension can be captured by a relatively simple analytical model, an
interesting direction for future research is to extend this model to
the mechanical action of proteins, such as RecA for example, on
supercoiled DNA.

\section*{Acknowledgments}
S. N. thanks Armand Ajdari (CNRS) for starting discussions on the
subject of electrostatic repulsion in DNA supercoiling.

\bibliography{DNAelectrostatics}

\end{document}